\documentclass[fleqn,usenatbib]{mnras}

\usepackage{newtxtext,newtxmath}

\usepackage[T1]{fontenc}
\usepackage{booktabs}
\DeclareRobustCommand{\VAN}[3]{#2}
\let\VANthebibliography\thebibliography
\def\thebibliography{\DeclareRobustCommand{\VAN}[3]{##3}\VANthebibliography}
\usepackage{xurl}

\usepackage{graphicx}	
\usepackage{amsmath}	
\usepackage{tabularx}
\usepackage[dvipsnames]{xcolor}
\newcommand{\tess}{TESS}

\title[TOI-2407\,b: a warm Neptune ]{TOI-2407\,b: a warm Neptune in the desert}

\author[C. Janó Muñoz et al.]{C. Janó Muñoz,$^{1}$\thanks{E-mail: cj467@cam.ac.uk}
M. J. Hooton,$^{1}$
P. P. Pedersen, $^{2}$
K. Barkaoui, $ ^{3, 4,19}$
B. V. Rackham, $ ^{4,5}$  
A. J. Burgasser, $ ^{6}$  
\newauthor
 F. J. Pozuelos, $ ^{7}$
K. G. Stassun, $ ^{8}$
D. Queloz, $ ^{1,2}$
A. H. M. J. Triaud, $ ^{9}$
C. Ziegler, $ ^{10}$
J. M. Almenara, $ ^{11, 12}$
\newauthor
  M. Timmermans, $ ^{9, 3}$ 
X. Bonfils, $ ^{11}$
K.A. Collins, $ ^{13}$
B. O. Demory, $ ^{14}$
G. Dransfield, $ ^{9, 20, 21}$
M. Ghachoui, $^{15, 3}$
\newauthor
M. Gillon, $ ^{3}$
E. Jehin, $^{16}$
A. W. Mann, $ ^{17}$
D. Sebastian, $ ^{9}$
S. Thompson, $ ^{1}$
J. D. Twicken, $^{18}$
J. de Wit, $^{4}$
\newauthor
S. Zúñiga-Fernández $ ^{3}$
\\
$^{1}$Cavendish Laboratory, JJ Thomson Avenue, Cambridge CB3 0HE, UK\\
$^{2}$ETH Zurich, Department of Physics, Wolfgang-Pauli-Strasse 2, 8093 Zurich, Switzerland\\
$^{3}$Astrobiology Research Unit, Université de Liège, 19C Allée du 6 Août, 4000 Liège, Belgium\\
$^{4}$Department of Earth, Atmospheric and Planetary Science, Massachusetts Institute of Technology, 77 Massachusetts Avenue, Cambridge, MA 02139, USA\\
$^{5}$Kavli Institute for Astrophysics and Space Research, Massachusetts
Institute of Technology, Cambridge, MA 02139, USA\\
$^{6}$Department of Astronomy \& Astrophysics, UC San Diego, 9500 Gilman Drive, La Jolla, CA 92093, USA \\
$^{7}$Instituto de Astrofísica de Andalucía (IAA-CSIC), Glorieta de la Astronomía s/n, 18008 Granada, Spain \\
$^{8}$Department of Physics \& Astronomy, Vanderbilt University, 6301
Stevenson Center Ln., Nashville, TN 37235, USA\\
$^{9}$School of Physics \& Astronomy, University of Birmingham,
Edgbaston, Birmingham B15 2TT, UK\\
$^{10}$Department of Physics, Engineering and Astronomy,  Stephen F. Austin State University,
1936 North Street
Nacogdoches, TX 75962, USA\\
$^{11}$Univ. Grenoble Alpes, CNRS, IPAG, 38000 Grenoble, France\\
$^{12}$Observatoire de Genève, Université de Genève, Chemin Pegasi, 51,
1290 Sauverny, Switzerland\\
$^{13}$Center for Astrophysics Harvard \& Smithsonian, 60 Garden Street, Cambridge, MA 02138, USA\\
$^{14}$Center for Space and Habitability, University of Bern, Gesellschaftsstrasse 6, 3012 Bern, Switzerland\\
$^{15}$Oukaimeden Observatory, High Energy Physics and Astrophysics
Laboratory, Faculty of sciences Semlalia, Cadi Ayyad University,
Marrakech, Morocco\\
$^{16}$STARInstitute, Université de Liège, Allée du 6 Août 19c, 4000 Liège, Belgium\\
$^{17}$Department of Physics and Astronomy, The University of North Carolina at Chapel Hill, Chapel Hill, NC 27599-3255, USA\\
$^{18}$SETI Institute, Mountain View, CA  94043, USA
NASA Ames Research Center, Moffett Field, CA  94035, USA\\
$^{19}$Instituto de Astrof\'isica de Canarias (IAC), Calle V\'ia L\'actea s/n, 38200, La Laguna, Tenerife, Spain\\
$^{20}$Department of Astrophysics, University of Oxford, Denys Wilkinson Building, Keble Road, Oxford OX1 3RH, UK\\
$^{21}$Magdalen College, University of Oxford, Oxford OX1 4AU, UK
}
\date{Accepted XXX. Received YYY; in original form ZZZ}

\pubyear{\the\year{}}

\begin{document}
\label{firstpage}
\pagerange{\pageref{firstpage}--\pageref{lastpage}}
\maketitle

\begin{abstract}
We present the validation of TOI-2407\,b, a warm Neptune-sized planet with a radius of $4.26\pm0.26$ R$_\oplus$, orbiting an early M-type star with a period of 2.7 days and an equilibrium temperature of $705\pm12$\,K. The planet was identified by TESS photometry and validated in this work through multi-wavelength ground-based follow-up observations. We include an observation with the novel CMOS-based infrared instrument SPIRIT at the SPECULOOS Southern Observatory. The high-precision transit data enabled by CMOS detectors underscore their potential for improving the detection and characterisation of exoplanets orbiting M-dwarfs, particularly in the infrared, where these stars emit most of their radiation. TOI-2407\,b lies within the boundaries of the period--radius Neptune desert, an apparent scarcity of Neptune-sized planets at short orbits. Further characterisation of TOI-2407\,b, such as radial velocity measurements, will refine its position within planetary demographic trends. This system also provides a comparison case for the well-studied Neptune-sized planet Gliese 436\,b, of similar radius, period and stellar type. Comparison studies could aid the understanding of the formation and evolution of Neptune-like planets around M-dwarfs.

\end{abstract}




\section{Introduction}

Space-based telescopes conducting transit surveys, particularly Kepler, launched in 2009 \citep[][]{Kepler}, and most recently TESS, launched in 2018 \citep[][]{Ricker2015}, have led to the discovery of thousands of exoplanets. This population contains a high abundance of small planets ($R_\text{p} < 4\,\text{R}_{\oplus}$) \citep[][]{Batalha2011, Dressing2013} across a wide range of parameter space without a direct Solar System analogue, such as super-Earths and sub-Neptunes. Statistical studies into the population of Neptune-sized planets have uncovered a scarcity of such planets in close-in orbits (2--4 day period) around their host stars, in a region of parameter space now called the Neptune desert \citep[][]{Szabo2011, Benitez2011, Mazeh2016}. The Neptune desert is likely to be caused by an interplay of formation and evolution mechanisms. The material availability of the inner disk may inhibit the formation of Neptune-sized planets, meaning planets in the Neptune desert have formed at an intermediate distance and migrated inwards \citep[][]{LeeChiang, DawsonJohnson2018}.Migration models suggest that the upper mass limit of the desert may be caused by tidal forces as the planet migrates inwards \citep[][]{Liu2013, Valsecchi2015}. At such close orbital distances, gaseous planets would be susceptible to tidal forces from the host star, leading to their disruption or significant mass loss through approach to the Roche limit, tidal heating or tidally-induced orbital decay. The lower mass limit of the desert is attributed to photoevaporation \citep[][]{OwenLai2018, Vissapragada2022}. High-energy radiation from the host star, particularly in the form of X-ray and extreme ultraviolet (XUV) flux during the first $\sim100$Myrs, strips the primordial H/He atmospheres of close-in planets over time \citep[][]{OwenWu}. This process is particularly relevant in scenarios of intense stellar irradiation. Planets with initial masses below the desert's lower limit are likely to lose their gaseous envelopes, leaving behind the cores.

The discovery of short-orbit sub-Neptunes, such as NGTS-4\,b \citep[][]{NGTS4} and more recently TOI-3261\,b \citep[][]{Nabbie2024}, have provided new information about the origin and boundaries of the Neptune desert. We expect extent and origin of the desert to vary with stellar type \citep[][]{MacDonald2019}, and M-dwarf systems offer a distinct test case due to their lower irradiation in short orbits. Characterisation of Neptune-like planets around M-dwarfs can then isolate the dominant mechanisms shaping the distribution of close-in Neptunes, particularly photoevaporation, bounds of the Neptune desert specific to cool stars.  

In this work, we validate a planet orbiting the M2-type star TOI-2407 (TIC 153078576) at a distance of 92 pc from the Sun. This target was initially observed with TESS and flagged as a TESS Object of Interest \citep[TOI;][]{TOI} in 2020. Follow-up ground-based photometric observations were taken with SPECULOOS \citep[][]{Delrez2018, Sebastian2021}, TRAPPIST-South \citep[][]{Jehin2011, Gillon2011}, LCOGT \citep[][]{LCOGT} and ExTrA \citep[][]{Bonfils2015}. 

In this work, the details on the observations and instruments used are shown in Section \ref{sec:obs}. Section \ref{sec:stellar_charac} showcases our stellar characterisation through spectroscopic (Section \ref{sec:spectroscopic}) and spectral energy distribution (Section \ref{sec:sed}) analysis; where we obtain the stellar properties we use throughout the work. We performed statistical and imaging-based validation of to rule out false-positive scenarios, explained in Section \ref{sec:validation}. Section \ref{sec:global_analysis} details our global analysis, with transit fitting on Section \ref{sec:transits} and search for additional planet candidates in the system on Section \ref{ss:sherlock}. Finally, we discuss our results and future prospects on Section \ref{sec:discuss}.

\section{OBSERVATIONS}
\label{sec:obs}

Given the TESS pixel scale of 21\arcsec, ground-based follow-ups are needed to rule out blended sources and confirm the transit event is on target. The target being relatively faint (See Table \ref{stellarpar}) disfavours spectroscopic follow-ups, so we collected photometric observations from ground-based instruments as part of the TESS Follow-up Observing Program (TFOP) \citep[][]{tfop}. The observations used to follow up on this target are summarised on Table \ref{tab:observations}, and the astrometric and photometric properties of the target are shown on Table \ref{stellarpar}.

\subsection{TESS}

The target was initially observed by TESS \citep[][]{Ricker2015} on sector 3 from UT 2018 September 21 to 2018 October 17 and 4 from 2018 October 19 to 2018 November 14, with 30-minute cadence. It was then re-observed with 2-minute cadence on sectors 30 and 31 from 2020 Sep 23 to 2020 Oct 20 and 2020 Oct 22 to 2020 Nov 18 respectively. The 2-minute cadence data were processed in the TESS Science Processing Operations Center \citep[SPOC;][]{SPOC} pipeline, and the 30-minute cadence data were processed in the TESS-SPOC High Level Science Product pipeline \citep[][]{TESS-SPOC}. The SPOC conducted a transit search of the 2-minute cadence light curve for sector 30 on UT 30 October 2020 with an adaptive, noise-compensating matched filter \citep[][]{Jenkins2002, Jenkins2010, Jenkins2020}, producing a Threshold Crossing Event with a 2.703 day period. An initial limb-darkened transit model was fitted \citep[][]{Li2019} and a suite of diagnostic tests were conducted to help assess the planetary nature of the signal \citep[][]{Twicken2018}. The transit signature passed all diagnostic tests presented in the SPOC Data Validation reports, and the source of the transit signal was localized within $2.71 \pm 2.63$ \arcsec of the target star. All TESS Input Catalog \citep[TIC;][]{Stassun_2018AJ_TESS_Catalog} objects other than the target star were excluded as potential sources of the transit signal. The TESS Science Office reviewed the vetting information and issued an alert for the candidate, TOI 2407.01, on UT 24 November 2020.

\begin{figure*}
    \centering
    \includegraphics[width=\textwidth]{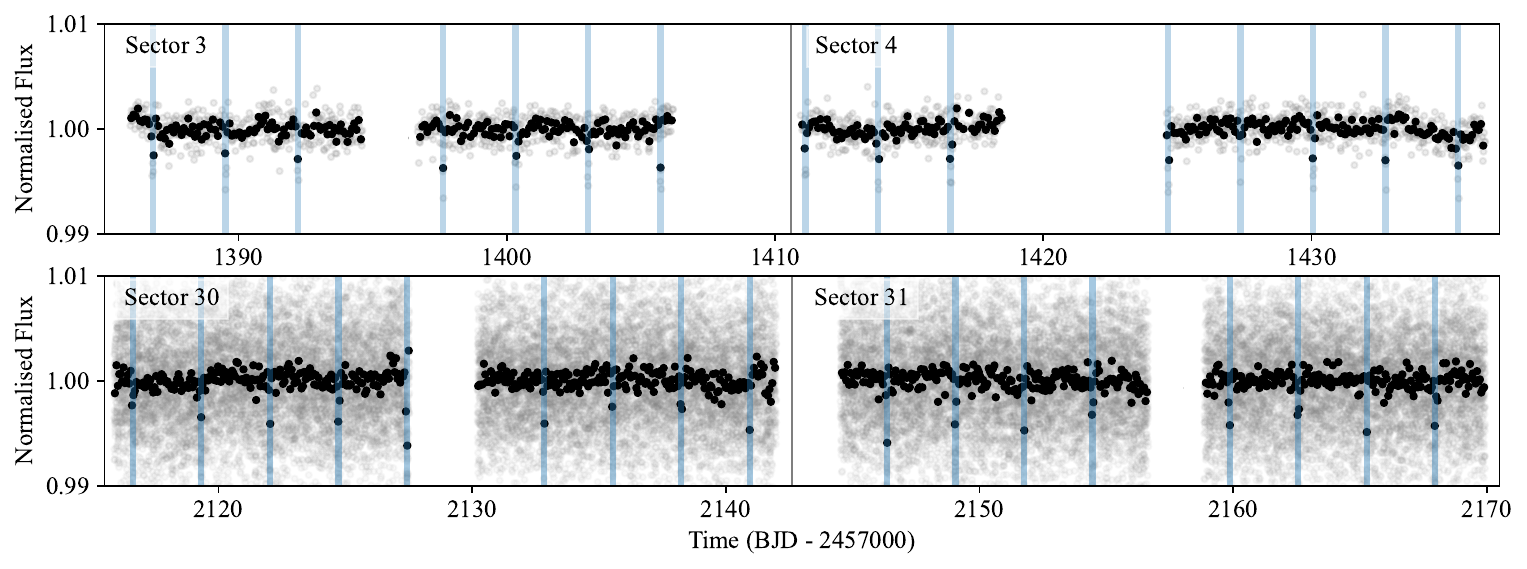}
    \caption{TESS lightcurves of TOI-2407 (TIC 15307857). Long-cadence sectors 3 and 4 are shown in the top panel and short-cadence sectors 30 and 31 in the bottom panel. Planetary transits are highlighted in blue. The binning showed in black corresponds to 1.5 hour bins.}
    \label{fig:tess_lcs}
\end{figure*}

The SPOC Presearch Data Conditioning Simple Aperture Photometry \citep[PDCSAP;][]{Stumpe2012, Stumpe2014, Smith2012} lightcurves, shown in Figure~\ref{fig:tess_lcs}, were obtained from the Mikulski Archive for Space Telescopes using the \texttt{lightkurve}\footnote{{\tt lightkurve:}~\url{https://github.com/lightkurve/lightkurve}} software \citep[][]{lightkurve}. The Simple Aperture Photometry (SAP) lightcurves were used to check for long-term stellar variability in the absence of high-resolution spectroscopic observations.

\subsection{Ground-based observations}

\subsubsection{SPECULOOS CCD photometry}

The SPECULOOS Southern Observatory (SSO) in Paranal Observatory, Chile, consists of four 1-m telescopes: Callisto, Europa, Ganymede and Io \citep[][]{Delrez2018}. All four are equipped with Andor-iKon 2\textit{k} $\times$ 2\textit{k} CCD cameras with a pixel scale of 0.35\arcsec, yielding a total field of view of $12\arcmin \times 12\arcmin$ \citep[][]{Burdanov2018}. These cameras are optimised for the near-infrared to observe ultra-cool M-dwarfs \citep[][]{Sebastian2021}. 

The target was observed by Europa on UT 21 December 2022 in the \textit{Sloan-$g'$} filter. It was re-observed on UT 24 October 2024 in the \textit{I+z} band. The data were processed with the SSO pipeline \citep[][]{Murray2020}, which performs bias, dark current and flat-field corrections during the data reduction process. The pipeline performs the photometry with varying aperture radii and selects the optimal one, in this case 6 pixels or 2.1\arcsec. It accounts for atmospheric and instrumental effects through differential photometry using a weighted sample of comparison stars. 

\subsubsection{TRAPPIST-South}
TRAPPIST-South  \citep[TRAnsiting Planets and PlanetesImals Small Telescope,][]{Jehin2011,Gillon2011} is a 60-cm Ritchey-Chr\'etien telescope installed at ESO-La Silla Observatory in Chile since 2010. The telescope is equipped with a 2\textit{k} $\times$ 2\textit{k} FLI Proline CCD with a  pixel-scale of 0.65\arcsec, with a resulting field of view of $22\arcmin \times 22\arcmin$.
Three full transits were observed with TRAPPIST-South on UT 13 December 2020, 1 January 2021 and 5 November 2022 in the $I+z$, $V$ and \textit{Sloan-$z'$} filters, with exposure times of 40, 120 and 90\,s, respectively. The data reduction and photometric extraction were performed using  {\tt PROSE}\footnote{{\tt Prose:} \url{https://github.com/lgrcia/prose}} pipeline \citep{prose_2022} and {\tt AstroImageJ}\footnote{{\tt AstroImageJ:}~\url{https://www.astro.louisville.edu/software/astroimagej}} \citep{Collins_2017}.

\subsubsection{ExTrA}

ExTrA \citep[Exoplanets in Transits and their Atmospheres,][]{Bonfils2015} is a low-resolution near-infrared (0.85 to 1.55~$\text{\textmu}$m) 
multi-object spectrograph fed by three 60-cm telescopes located at La Silla Observatory in Chile. Nine transits of TOI-2407.01 were observed using one, two or three of the ExTrA telescopes, and 8$\arcsec$ diameter aperture fibres and the 
low-resolution mode ($R$$\sim$20) of the spectrograph, with an exposure time of 60\,s. Five fibres are positioned in the focal plane
of each telescope to select light from the target and four comparison stars. The resulting ExTrA data were analyzed using custom data reduction software.

\subsubsection{LCO-CTIO}
To confirm the event on target and check the field for possible nearby eclipsing binaries, TOI-2407 was observed on UT 6 December 2020 with the Las Cumbres Observatory Global Telescope (LCOGT; \citealt{Brown:2013})  1.0-m network node at Cerro Tololo Inter-American Observatory (CTIO). Observations were in the \textit{Sloan-$i'$} band and consisted of 82\,s exposures. Science images were calibrated using the standard LCOGT {\tt BANZAI} pipeline \citep{McCully:2018}, and  photometric measurements were extracted using the {\tt AstroImageJ} software \citep{Collins:2017}. 

\subsubsection{SPECULOOS CMOS photometry with SPIRIT}
\label{sec:cmos_photometry}
One of the SPECULOOS transits was obtained with the complementary metal-oxide-semiconductor (CMOS) infra-red detector SPIRIT \citep[][]{SPIRIT}. CMOS detectors \citep[][]{CMOS} differ from CCDs \citep[][]{CCD} in structural design and operation. CCD detectors work by transferring pixel charges across the sensor array to a readout node, where they are then converted to voltage and processed. This sequential transfer offers high image quality with low noise levels but slow readout. In contrast, for CMOS sensors, each pixel is equipped with its own amplifier. This allows parallel readout across the entire array, meaning significantly faster readout speeds. The fast readout provides an advantage for high-cadence time-series observations. Recent use of CMOS sensors in astronomy show these instruments can achieve photometric accuracies and noise levels comparable to those of their CCD counterparts \citep[][]{Alarcon_2023}. 

SPIRIT is a 1280 $\times$ 1024 near-infrared InGaAs CMOS detector installed on the Callisto telescope in the SSO \citep[][]{SPIRIT}. It is equipped with a custom wide-pass filter, \textit{zYJ}, that operates at 0.81--1.33\,$\text{\textmu}$m. Given that SPIRIT operates in the near-infrared, its performance is expected to be advantageous for ultracool dwarfs observations beyond 1.1\,$\text{\textmu}$m. This is a regime that traditional silicon-based detectors, such as CCDs, cannot reach due to silicon's intrinsic wavelength cut-off.

TOI-2407 was observed by SPIRIT on UT 21 December 2022 with 10\,s exposures. The data were processed with a modified version of the \cite{Murray2020} SPECULOOS pipeline, optimised for CMOS and for the near infrared. In CMOS detectors, the current flowing through each pixel's readout integrated circuit (ROIC) produces infrared photons, which causes a glow in the image \citep[][]{ROIC}. This is mitigated by using dark current frames with exposure times matching that of the science images. CMOS detectors also display scattered bad pixels across the sensor, less compactly distributed than CCDs \citep[][]{badpixels}. This is addressed in the pipeline through nightly bad pixel mapping. The effect of precipitable water vapour (PWV) is minimised by the use of the custom-made \textit{zYJ} filter. SPIRIT produces high time-cadence data with a read-out time of 0.1\,s.
\begin{table*}
\caption{Summary of ground-based observations. SSO refers to the SPECULOOS Southern Observatory, where the zYJ filter on the Callisto telescope is the SPIRIT infrared detector. The de-trending parameters chosen based on the Bayesian Information Criterion (BIC) are shown on the right-most column for each instrument and filter.}
    \begin{tabularx}
    {\linewidth}{l>{\centering}X>{\centering\arraybackslash}cccc}
    \toprule
        \textbf{Instrument} & \textbf{Filter} & \textbf{Aperture (pix)} & \textbf{Pixel size ($\arcsec$)} & \textbf{Exposure (s)} & \textbf{De-trending parameters} \\ \midrule 
        SSO Europa & \textit{Sloan-$g'$} & 6& 0.35 & 90.0 &Time, airmass, FWHM, y displacement\\ 
        SSO Io & $I+z$ & 6& 0.35 & 10.0 & Time, airmass\\ 
        SSO Callisto & \textit{zYJ} & 4& 0.309 & 10.0 & Time, background \\ \midrule
        TRAPPIST South & $V$ & 8 & 0.64 & 120.0 & Time, background, airmass\\ 
        TRAPPIST South & \textit{Sloan-$z'$} & 5 & 0.64 & 90.0 & Time, airmass, y displacement, meridian offset\\ 
        TRAPPIST South & $I+z$ & 5 & 0.64 & 40.0 & Time, meridian offset\\ 
        \midrule
        ExTrA & 1.2\,$\text{\textmu}$m & - & - & 60.0 & Time, airmass, background, x and y displacement\\ 
        \midrule
        LCO-CTIO & \textit{Sloan-$i'$} & 13 & 0.39 & 82.0 & Time\\ 
        \hline
    \end{tabularx}
    \label{tab:observations}
\end{table*}

\section{Stellar properties of TOI-2407}
\label{sec:stellar_charac}
\begin{table}
\centering
\caption{Astrometry and photometry of TOI-2407. 
	{\bf (1)} Gaia EDR3 \citealt{Gaia_Collaboration_2021AandA}; 
	{\bf (2)} TESS Input Catalog \citealt{Stassun_2018AJ_TESS_Catalog}; 
	{\bf (3)} UCAC4 \citealt{Zacharias_2012yCat.1322};
	{\bf (4)} 2MASS \citealt{Skrutskie_2006AJ_2MASS};
	{\bf (5)} WISE \citealt{Cutri_2014yCat.2328}.
	}
\begin{tabularx}{\linewidth}{l>{\centering}X>{\centering\arraybackslash}c}
\toprule
			\multicolumn{3}{l}{Target designations}   \\
			\midrule
\multicolumn{2}{l}{TOI-2407}  \\
\multicolumn{2}{l}{TIC 153078576}  \\
\multicolumn{2}{l}{GAIA DR3 4847001302277419904 } \\\multicolumn{2}{l}{2MASS J03185915-4627273}   \\
			 \midrule 
   			\textbf{Parameter} & \textbf{Value} &  \textbf{Source}   \\
            \midrule
			\multicolumn{3}{l}{Parallax and distance}   \\\midrule
			RA (J2000)     & 03:18:59.17 &  (1) \\
			Dec (J2000)    & -46:27:28.89 &  (1)\\
			$\pi$ (mas) & $10.845 \pm 0.012$ &   (1)\\
                $\mu_{\text{RA}}$ (mas\,yr$^{-1}$) & $10.868 \pm 0.012$ & (1) \\
                $\mu_{\text{Dec}}$ (mas\,yr$^{-1}$) & $-98.287 \pm 0.013$ & (1) \\
			Distance (pc)  & $92.2 \pm 0.1$ & (1)\\
			\midrule
			\multicolumn{3}{l}{Photometric properties}   \\
            \midrule
			TESS$_{\rm mag}$           &  $12.526 \pm 0.007$  & (2)  \\
			$V_{\rm mag}$      & $ 14.68 \pm 0.11$  & (3) \\
            $B_{\rm mag}$       & $15.97 \pm 0.05$  & (3) \\
			$J_{\rm mag}$      & $11.13 \pm 0.02$  &  (4) \\
			$H_{\rm mag}$      & $10.56 \pm 0.03$  &    (4) \\
			$K_{\rm mag}$      & $10.32 \pm 0.02$  &   (4) \\			
			$G_{\rm mag}$   & $13.6328 \pm 0.0004$  &  (1)  \\
			$W1_{\rm mag}$     & $10.22 \pm 0.02$ &  (5) \\
			$W2_{\rm mag}$      & $10.16 \pm 0.02$  &   (5) \\
			$W3_{\rm mag}$     & $10.01 \pm  0.04$  &   (5)\\
            $W4_{\rm mag}$      & $ 8.57$  &   (5)\\
            \hline
	\end{tabularx}
    \label{stellarpar}
\end{table}

\subsection{Spectroscopic analysis}
\label{sec:spectroscopic}

\subsubsection{SOAR/TripleSpec4.1}

We collected a medium-resolution near-infrared spectrum of TOI-2407 on UT 19 September 2024 with the TripleSpec4.1 spectrograph \citep{Schlawin2014} on the Southern Astrophysical Research (SOAR) 4.1\,m telescope using the Astronomical Event Observatory Network (AEON) queue. The fixed 1\farcs{1} slit of the instrument provides $R{\sim}3500$ spectra covering 0.80–2.47\,$\text{\textmu}$m. Nodding in an ABBA pattern, we gathered four exposures of 120.8\,s, and four 13.8\,s exposures of the A0\,V standard HD\,21638 ($V = 9.57$). We reduced the data using Spextool v4.1\footnote{{TripleSpec reduction}:~\url{https://noirlab.edu/science/observing-noirlab/observing-ctio/observing-soar/data-reduction/triplespec-data}.} \citep{Cushing2004}, 
modified for TripleSpec4.1 data, using the standard set of flat field and arc lamp exposures collected the previous afternoon. The final spectrum has a median S/R per pixel of 49 and an average of 1.4\,pixels per resolution element.

\begin{figure}
    \centering
    \includegraphics[width=0.99\linewidth]{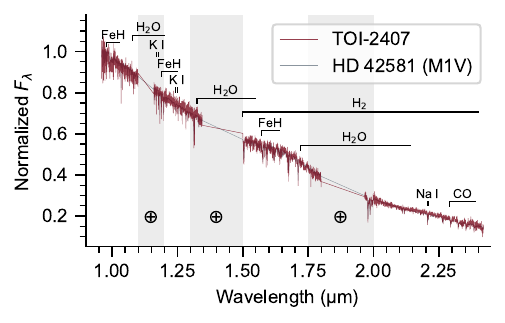}
    \caption{TripleSpec4.1 spectrum of TOI-2407 (red), alongside the IRTF/SpeX spectrum of M1 standard HD\,42581 (grey) for comparison.
Regions of strong telluric absorption are shaded, and prominent spectral features of M dwarfs are highlighted.
The TripleSpec4.1 spectrum has a higher resolving power than the SpeX spectrum ($R{\sim}3500$ vs.\ $R{\sim}2000$), giving it a more jagged appearance.}
    \label{fig:triplespec}
\end{figure}

Fig.\,\ref{fig:triplespec} shows the TripleSpec4.1 spectrum of TOI-2407.
Comparing it to single-star spectral standards in the IRTF Spectral Library \citep{Cushing2005, Rayner2009} using the SpeX Prism Library Analysis Toolkit \citep[SPLAT;][]{splat}, we find the best match to the M1\,V standard HD\,42581 and thus adopt an infrared spectral type of M1.0 $\pm$ 0.5.
We also used SPLAT to measure the H2O--K2 index \citep{Rojas-Ayala2012} and the equivalent widths of the $K$-band Na\,\textsc{i} and Ca\,\textsc{i} doublets.
Using the empirical relation between these features and stellar metallicity \citep{Mann2013}, and propagating uncertainties with a Monte Carlo approach \citep[see, e.g.,][]{Barkaoui2024, Ghachoui2024}, we estimate an iron abundance of $\mathrm{[Fe/H]} = +0.20 \pm 0.16$ for TOI-2407.

\subsubsection{Magellan/MagE}

TOI-2407 was also observed on UT 6 October 2022 with the Magellan Echellette spectrograph (MagE; \citealt{2008SPIE.7014E.169M}) on the 6.5-m Magellan Baade Telescope.
Conditions were clear with seeing of $0\farcs9$.
We used the 0$\farcs$7$\times$10$\arcsec$ slit to obtain resolution $\lambda/\Delta\lambda$ $\approx$ 6000 over 4000--9000\,{\AA}, and obtained two 100-s exposures at an average airmass of 1.05.
We also observed the spectrophotometric calibrator Feige\,110 for flux calibration \citep{1992PASP..104..533H,1994PASP..106..566H}, and obtained bias, ThAr arc lamp, Xe flash lamp, and incandescent lamp exposures at the start of the night for wavelength and flux calibration. 
We did not observe a telluric absorption calibrator for these observations, hence strong telluric features from oxygen and water remain in the spectrum.
Data were reduced using PypeIt \citep{pypeit:zenodo, pypeit:joss_arXiv, pypeit:joss_pub} using standard settings. The final calibrated spectrum has a median 
S/N $\approx$ 60 at 5450~{\AA}
and S/N $\approx$ 105 at 7450~{\AA}. 

\begin{figure}
    \centering
    \includegraphics[width=0.99\columnwidth, keepaspectratio]{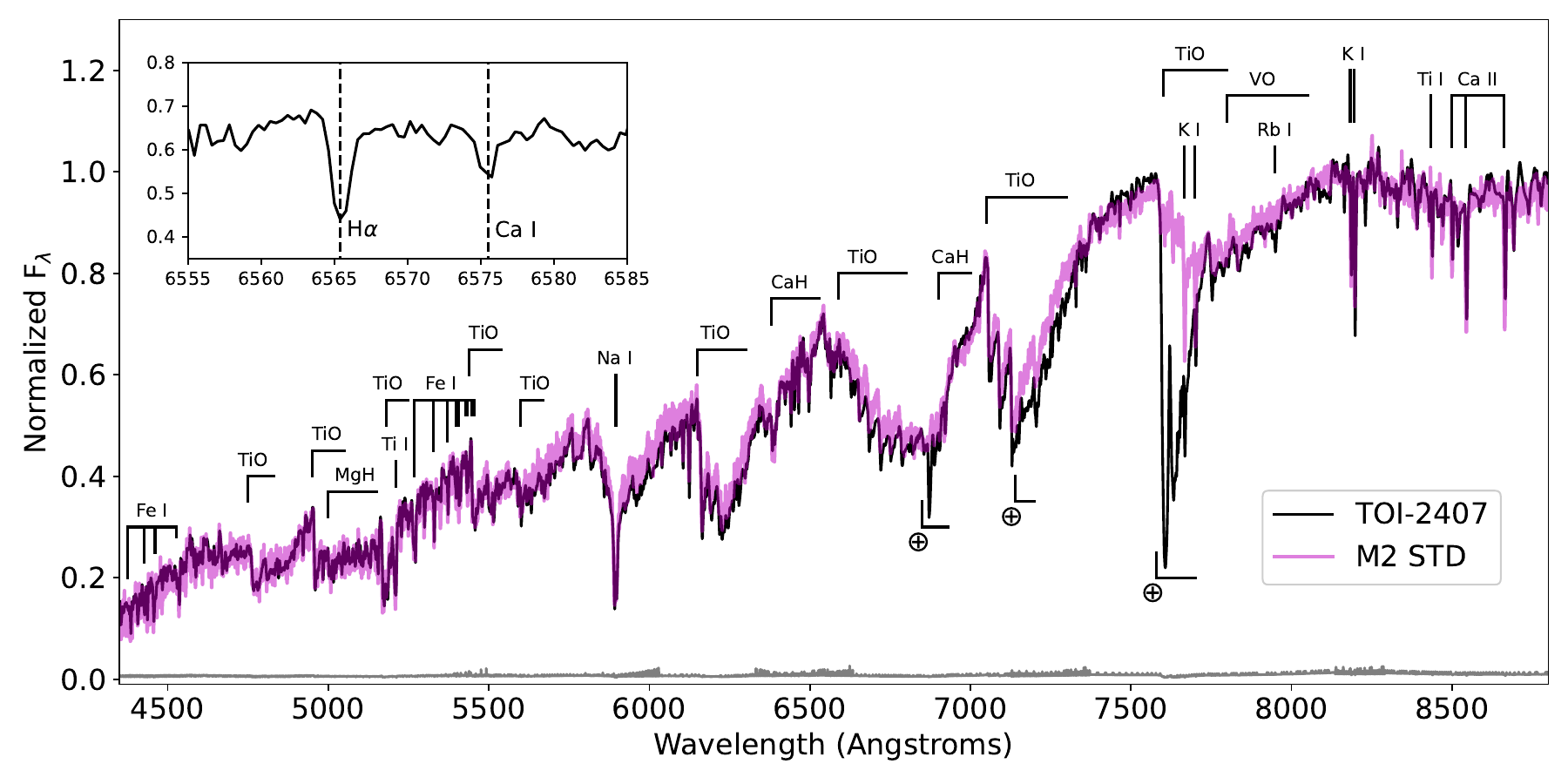}
    \caption{
        Baade/MagE spectrum of TOI-2407 compared to the M2 SDSS template from \citet[][ magenta line]{2007AJ....133..531B}.
        Data are normalized at 7400~{\AA}, and key spectral features across the 4300--8100~{\AA} region are labelled, including the location of uncorrected telluric oxygen and water bands ($\oplus$).
        The inset box show the 6555--6585~{\AA} region with H$\alpha$ and Ca~I absorption features labelled.       }
    \label{fig:mage}
\end{figure}

Fig.\,\ref{fig:mage} displays the MagE spectrum of TOI-2407 over the 4200--8800~{\AA} range. We used analysis tools in the the \texttt{kastredux} package\footnote{\texttt{kastredux}:~\url{https://github.com/aburgasser/kastredux}.} to characterize the spectrum.
Comparison to M dwarf spectral templates from \citet{2007AJ....133..531B} finds a best match to a slightly later spectral type, M2, which is consistent with index-based classifications from 
\citet[][ M2.0$\pm$0.3]{1995AJ....110.1838R}; 
\citet[][ M1.5$\pm$0.4]{1997AJ....113..806G}; and 
\citet[][ M2.5$\pm$0.7]{2003AJ....125.1598L}.
H$\alpha$ is observed in absorption with an equivalent width EW = 0.30$\pm$0.03~${\AA}$, which implies an age $\gtrsim$1.2~Gyr \citep{2008AJ....135..785W}, consistent with the large space velocity of the system relative to the Sun 
$(U,V,W)$ = (27.11$\pm$0.16, $-$52.2$\pm$0.7, $-$29.5$\pm$1.0)~km$/$s \citep{2023A&A...674A...1G}.
We measure the metallicity index $\zeta$ from CaH and TiO absorption features near 7000~{\AA} \citep{2007ApJ...669.1235L,2013AJ....145...52M}, finding a value of $\zeta$ = 1.24$\pm$0.02 that suggests a supersolar metallicity ([Fe/H] $\gtrsim$ $+0.2$;  \citealt{2013AJ....145...52M}), consistent with our near-infrared spectral analysis.
Low-resolution spectral measurements from Gaia DR3 also indicate a slightly supersolar metallicity, [Fe/H] = 0.117$\pm$0.003 \citep{2023A&A...674A..27A}.

\subsection{Spectral energy distribution}
\label{sec:sed}

\begin{figure}
    \centering
    \includegraphics[width=0.99\linewidth,trim=80 70 50 60,clip]{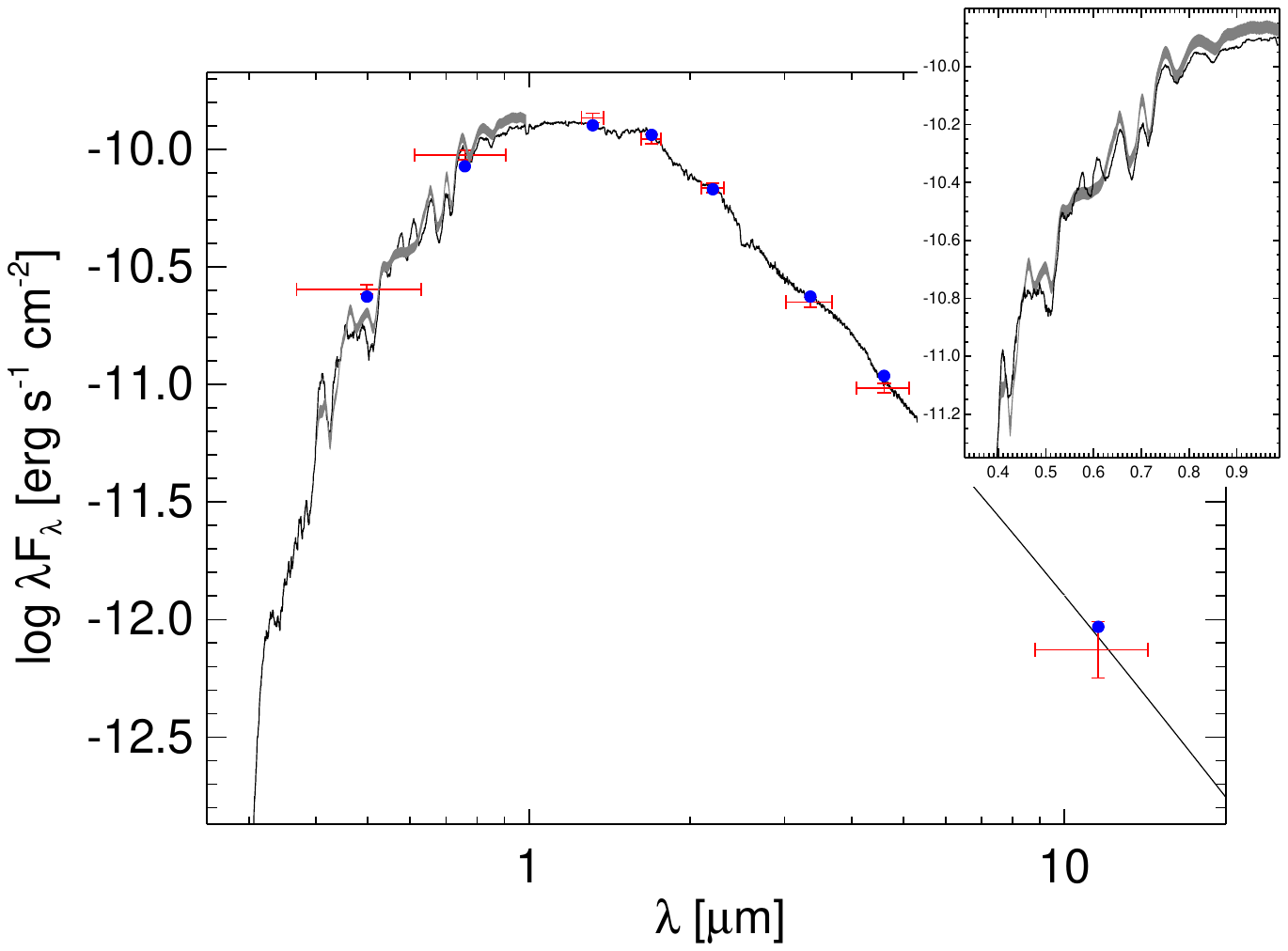}
    \caption{Spectral energy distribution (SED) of the host star TOI-2407. Red crosses represent the observed photometric measurements, where the horizontal bars represent the effective width of the passband. Blue points are the modelled fluxes from the best-fit Kurucz atmosphere model (black). The inset panel shows the absolute flux-calibrated {\it Gaia\/} spectrophotometry as a gray swathe.}
    \label{fig:sed}
\end{figure}

As an independent determination of the basic stellar parameters, we performed an analysis of the broadband spectral energy distribution (SED) of the star together with the Gaia DR3 parallax \citep[with no systematic offset applied; see, e.g.,][]{StassunTorres:2021}, in order to determine an empirical measurement of the stellar radius, following the procedures described in \citet{Stassun:2016,Stassun:2017,Stassun:2018}. We pulled the $JHK_S$ magnitudes from 2MASS, the W1--W3 magnitudes from WISE, and the $G G_{\rm BP} G_{\rm RP}$ magnitudes from Gaia. We also utilized the absolute flux-calibrated Gaia spectrophotometry. Together, the available photometry spans the full stellar SED over the wavelength range 0.4--10~$\text{\textmu}$m (see Figure~\ref{fig:sed}).

We performed a fit using PHOENIX stellar atmosphere models \citep{Husser:2013}, with the principal parameters being the effective temperature ($T_{\rm eff}$), surface gravity ($\log g$) and metallicity ([Fe/H]), adopted from the spectroscopic analysis. We limited the extinction, $A_V$, to the maximum line-of-sight value from the dust maps of \citet{Schlegel:1998}. The resulting fit, shown in Figure~\ref{fig:sed}, has a reduced $\chi^2$ of 2.2 and $A_V = 0.02 \pm 0.02$. Integrating the (unreddened) model SED gives the bolometric flux at Earth, $F_{\rm bol} = 1.692 \pm 0.059 \times 10^{-10}$ erg~s$^{-1}$~cm$^{-2}$. Taking the $F_{\rm bol}$ and $T_{\rm eff}$ together with the Gaia parallax, gives the stellar radius, $R_\star = 0.567 \pm 0.034$~R$_\odot$. In addition, we can estimate the stellar mass from the empirical relations of \citet{Mann:2019}, giving $M_\star = 0.548 \pm 0.016$~M$_\odot$. We use this mass and radius in our analysis. 

\section{Validation of the planet}
\label{sec:validation}

\subsection{Statistical validation}

We considered astrophysical false-positive scenarios through statistical validation using \texttt{TRICERATOPS}\footnote{{\tt TRICERATOPS:}~\url{https://github.com/stevengiacalone/triceratops}}  \citep[][]{triceratops, triceratops_code}. This software uses a Bayesian framework to calculate the overall false-positive probability (FPP) and the nearby false-positive probability (NFPP), which is the probability that the transit feature is caused by a nearby star in the field. For a false positive scenario to be ruled out, the values must be NFPP $ < 10^{-3}$ and FPP $< 0.015$. Using the 4 TESS sectors, we obtained an FPP of 0.005 $\pm$ 0.004 and NFPP $< 10^{-4}$. Although this statistically rules out the transit feature being produced by resolved stars, we use archival and high-resolution imaging to confirm it is not caused by unresolved sources.   

\subsection{High-resolution and archival imaging}
\begin{figure}
    \centering
    \includegraphics[width=0.5\textwidth]{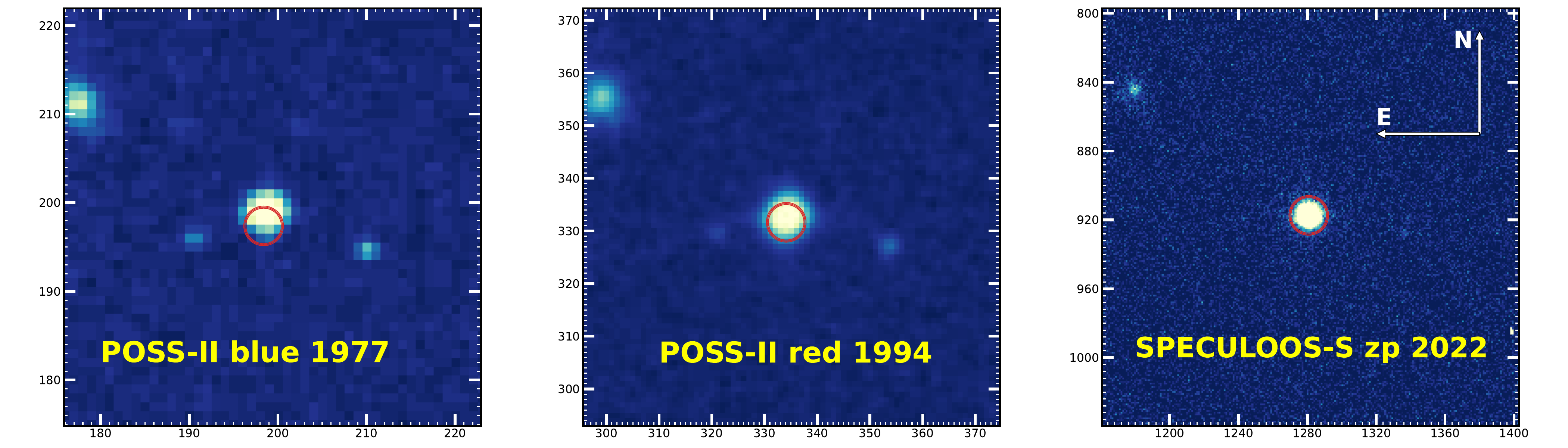}
    \caption{Archival images of TOI-2407 and its field taken on 1977, 1994 and 2022 by DSS/POSS-II and SPECULOOS South. From left to right taken with the POSS-II blue plate, POSS-II infrared plate and SPECULOOS \textit{I+z} band. The red circle shows the current position of the target.}
    \label{fig:archival}
\end{figure}

It is necessary to search for nearby sources that can contaminate the TESS photometry, resulting in an underestimated planetary radius, or be the source of astrophysical false-positives, such as background eclipsing binaries. We used archival images from POSS-II/DSS \citep[][]{possii} and 2MASS \citep[][]{2MASS} to find nearby stars that could be currently unresolved by TESS. From the images we discarded any nearby sources closer than 30$\arcsec$ to the target with a $J$-magnitude upper limit of 22.5, shown in Figure \ref{fig:archival}. However, TOI-2407 has a low proper motion of $99$\,mas\,yr$^{-1}$ \citep[][]{GaiaDR3}, meaning a total motion of $\sim4\arcsec$ since the first POSS-II observation of the target in 1977. In this case, discarding the presence of contaminating background objects through archival imaging is not enough. We thus obtained high-resolution imaging to further constrain the presence of a companion object.

\begin{figure}
    \centering\includegraphics[width=0.4\textwidth]{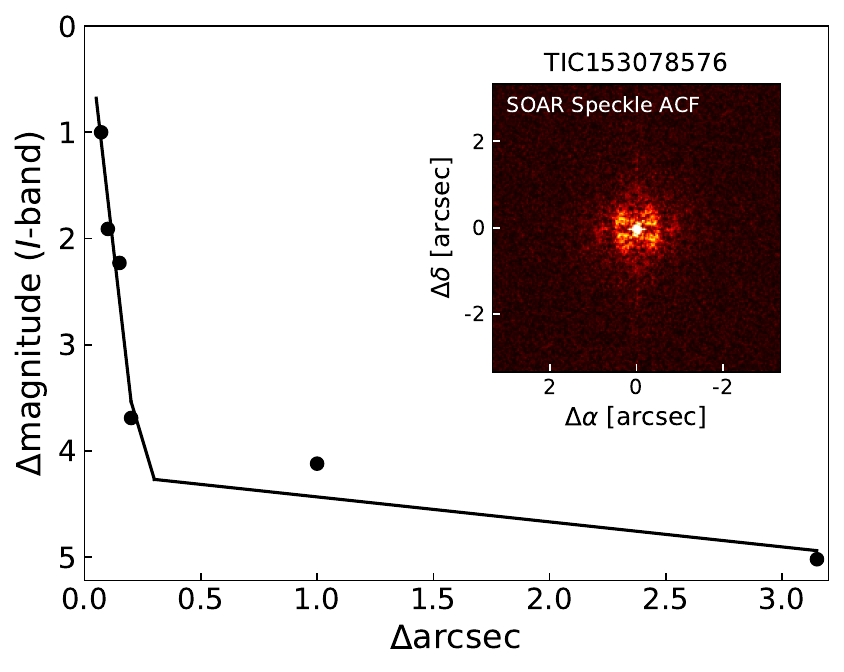}
    \caption{Speckle autocorrelation function (red; top-right corner) and detection curve (black) of the SOAR Cousins-I observation, observed on UT 3 December 2020.}
    \label{fig:SOARimaging}
\end{figure}

We searched for stellar companions to TOI-2407 with speckle imaging on the 4.1-m Southern Astrophysical Research (SOAR) telescope \citep[][]{SOAR} on 3 December 2020 UT, observing in the \textit{Cousins I} band, a visible bandpass similar to TESS. This observation was sensitive to a 4.2-magnitude fainter star at an angular distance of 1$\arcsec$ from the target. More details of the observations within the SOAR TESS survey are available in \cite{Ziegler2020}. The 5$\sigma$ detection sensitivity and speckle autocorrelation function of the observations are shown in Figure \ref{fig:SOARimaging}. No nearby stars were detected within 3\arcsec of TOI-2407 in the SOAR observations.

\section{Transit analysis}
\label{sec:global_analysis}
\subsection{Transit fitting}
\label{sec:transits}
The lightcurves were analysed globally with a transit model, a stellar activity correction, and a baseline model on the ground-based data to account for atmospheric effects and instrumental systematics.

The transit model was calculated with \texttt{batman}\footnote{{\tt batman:}~\url{https://lkreidberg.github.io/batman}} \citep[][]{batman}. This Python packages calculates lightcurves from input parameters, in our case: orbital period $P$, mid-transit time $t_0$, planet-star radius ratio, impact parameter $b$, eccentricity and argument of periastron as $\sqrt{e} \cos\omega$ and $\sqrt{e} \sin\omega$, stellar density $\rho_*$, and quadratic limb darkening coefficients.

\begin{figure}
    \centering
    \includegraphics[width=0.95\linewidth]{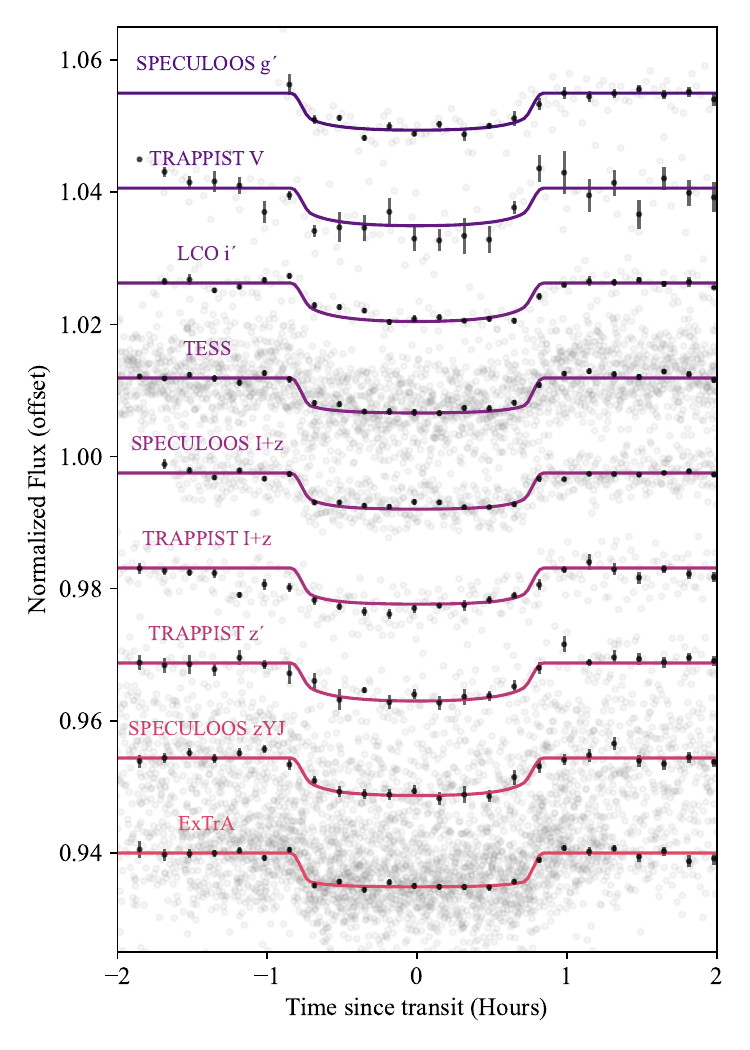}
    \caption{Phase-folded transits with 10 minute binning on a flux offset, from bluest to reddest wavelength (top to bottom). The fitted transit model is shown on top. The SPIRIT fit is labeled SPECULOOS zYJ. Additionally, the SPECULOOS \textit{I+z} data show evidence of a stellar spot crossing.}
    \label{fig:fits}
\end{figure}

A baseline model was applied independently to each ground-based lightcurve. These baselines were built as linear combinations of observable parameters such as airmass, sky background, full-width-half-maximum (FWHM) of the PSF, and displacement on the x and y axes of the detector. The baseline parameters used for each night were determined with the Bayesian Information Criterion \citep[BIC;][]{BIC} and are summarised in Table \ref{tab:observations}.

The stellar activity correction was applied to the TESS data using a Gaussian Process (GP) with \texttt{celerite} \citep[][]{celerite}. This  accounts for stellar variation that is unexplained through physical modelling. The kernel chosen was a semi-periodic stochastically-driven damped harmonic oscillation with a periodicity term $\omega_0$, a noise amplitude term $S_0$, and a quality factor $Q$ which was fixed to $1/\sqrt{2}$. 

\begin{table*}
\centering
\caption{Stellar and planetary parameters excluding limb-darkening coefficients, which are listed in Appendix \ref{limbdarkening}. The stellar parameters, except the density, were obtained through spectral and SED analysis shown in Section \ref{sec:stellar_charac}. The density is consistent with the mass and radius. The planet parameters with empty prior columns are fit parameters obtained without informative priors and derived parameters.}
\renewcommand{\arraystretch}{1.2} 
\begin{tabularx}{\textwidth}{l>{\centering}X>{\centering\arraybackslash}cc}
\toprule
\textbf{Parameter} & \textbf{Value} & \textbf{Priors} \\ \midrule
\multicolumn{3}{l}{Stellar parameters} \\ \midrule
Stellar density (g\,cm$^{-3}$) & \(4.85^{+0.28}_{-0.29}\,\) & \(\mathcal{N}(4.24, 0.77)\) \\
$T_{\text{eff}}$ (K) & \(3530 \pm 100\,\) & -\\
$\text{[Fe/H]}$ & \(0.21 \pm 0.16\,\) & -\\
$R_*$ (\(\text{R}_\odot\)) & $0.567 \pm 0.034$ & -\\
$M_*$ (\(\text{M}_\odot\)) & $0.548 \pm 0.016$ & -\\
$L_*$ (\(\text{L}_\odot\)) & $0.0448 \pm 0.0016$ & -\\
Bolometric flux $F_{\text{bol}}$ (erg\,s$^{-1}$\,cm$^{-2}$) & $(1.692 \pm 0.059)\times10^{-10}$ & -
\\\midrule
\multicolumn{3}{l}{Planet parameters} \\ \midrule
Period (days) & \(2.702969 \pm 0.000001\) & \(\mathcal{N}(2.703, 0.05)\) \\
Mid-transit time \(t_0\) (BJD - 2450000) & \(9149.0561 \pm 0.0003\) & \(\mathcal{U}(9149.05, 9149.06)\) \\
Transit duration (hours) & \(1.70^{+0.04}_{-0.09}\,\) & -\\
\( R_p / R_* \) & \(0.0689^{+0.0008}_{-0.0009}\,\) & \(\mathcal{U}(0.01, 0.1)\) \\
Radius (\(\text{R}_\oplus\)) & \(4.26 \pm 0.26\) & - \\
Transit impact parameter \(b\) & \(0.25^{+0.09}_{-0.11}\,\) & \(\mathcal{U}(0, 1)\) \\
Orbital inclination $i$ (deg) & \(88.83^{+0.55}_{-0.41}\,\) & - \\
\(\sqrt{e}\cos{\omega}\) & \(0.19^{+0.17}_{-0.13}\,\) & \(\mathcal{U}(-1, 1)\) \\
\(\sqrt{e}\sin{\omega}\) & \(0.16^{+0.07}_{-0.10}\,\) & \(\mathcal{U}(-1, 1)\) \\
Eccentricity \(e\) & \(0.08^{+0.07}_{-0.04}\,\) & - \\
Argument of periapsis \(\omega\) (deg) & \(42^{+32}_{-30}\,\) & - \\
Semi-major axis (AU) & \(0.033 \pm 0.001\) & - \\
Stellar irradiation (\(\text{S}_\oplus\)) & \(41 \pm 4\) & - \\
Equilibrium temperature (K) & \(705 \pm 12\) & - \\ 
\hline
\end{tabularx}
\label{tab:results_general}
\end{table*}

The models were then fit to the lightcurves with \texttt{emcee} \citep[][]{emcee}, a software implementation of an affine invariant Markov chain Monte Carlo (MCMC) ensemble sampler \citep[][]{MCMC}. We used uniform priors on all free parameters except $P$ and $\rho_*$. For the period, a Gaussian prior was used based on an initial period estimate through a periodogram of the TESS data. The stellar density prior was also Gaussian, determined by the mass and radius values obtained in Section \ref{sec:sed}. 
The eccentricity $e$ and the argument of the periastron $\omega$ were fitted together by sampling $\sqrt{e} \cos\omega$ and $\sqrt{e} \sin\omega$ without informative priors, to avoid biasing the posterior towards non-zero eccentricities. For the stellar activity GP kernel, we used wide uniform priors on log($\omega_0$) and log($S_0$).

The stellar limb darkening was modelled with the quadratic limb darkening described in \cite{Mandel2002}, which fits two quadratic parameters: $u_1$ and $u_2$. In our MCMC, to avoid unphysical combinations of these parameters, we sample the transformed triangular quadratic limb darkening parameters from \cite{Kipping2013_limb},
\begin{align}
q_1 &= (u_1 + u_2)^2 \\
q_2 &= \frac{u_1}{2(u_1 + u_2)}.
\end{align}
As shown in Appendix \ref{limbdarkening}, we decided to use wide uniform priors on both parameters following the evidence of deviation between the empirical and tabulated limb-darkening parameters shown in \cite{Patel2022}. Since the limb darkening is wavelength dependent, we sampled this parameter separately for lightcurves of different filters.

We ran the MCMC with 100 walkers for 5000 steps to sample the posterior distribution of each fit parameter and calculate its uncertainties. Our results are shown in Table \ref{tab:results_general} and Figure \ref{fig:fits}. Chromaticity checks confirmed the consistency of the transit depths at each wavelength.

The SPIRIT light curve is displayed in Figure \ref{fig:fits} (second from the bottom). Due to its infrared-sensitive CMOS architecture, SPIRIT exhibits a higher thermal noise contribution compared to the other instruments, leading to an increased root-mean-square (RMS) scatter in the light curve. However, limb darkening effects decrease at longer wavelengths, reducing degeneracies in transit depth measurements, which improves the precision of planetary radius estimates.

{\subsection{Search for additional candidates}\label{ss:sherlock}}

We looked for evidence of additional planets in the system by searching the TESS data for possible missed transits, performing injection recovery to calculate the upper limits of a non-detected planet, and looking for transit timing variations (TTVs).  

For checking the presence of other transits in the TESS data, we used the \texttt{SHERLOCK}\footnote{{\tt SHERLOCK:}~\url{https://github.com/franpoz/SHERLOCK}} software \citep[][]{sherlock2, sherlock}. \texttt{SHERLOCK} is an end-to-end pipeline that allows the users to explore the data from space-based missions to search for planetary candidates. It can be used to recover alerted candidates by the automatic pipelines such as SPOC \citep[][]{SPOC} in our case, and to search for candidates that remain unnoticed due to detection thresholds, lack of data exploration, or poor photometric quality. To this end, \texttt{SHERLOCK} has seven different modules to (1) acquire and prepare the light curves from their repositories, (2) search for planetary candidates, (3) vet the interesting signals, (4) perform a statistical validation, (5) carry out planetary system stability tests, (6) model the signals to refine their ephemeris, and (7) compute the observational windows from ground-based observatories to trigger a follow-up campaign.

We ran \texttt{SHERLOCK} on all four sectors covering possible periods from 0.5 to 33 days with 3 runs each. TOI-2407\,b was recovered on the initial run. One additional candidate was flagged, but only observed in three of the four sectors. This candidate had a depth of $1.4 \pm 0.4$ ppt and period of 26.3 days with a signal-to-noise ratio of 4.4. The S/R is below 5 and the candidate is not statistically significant so we did not explore it further in this work.

\begin{figure}
    \centering
    \includegraphics[width=0.4\textwidth]{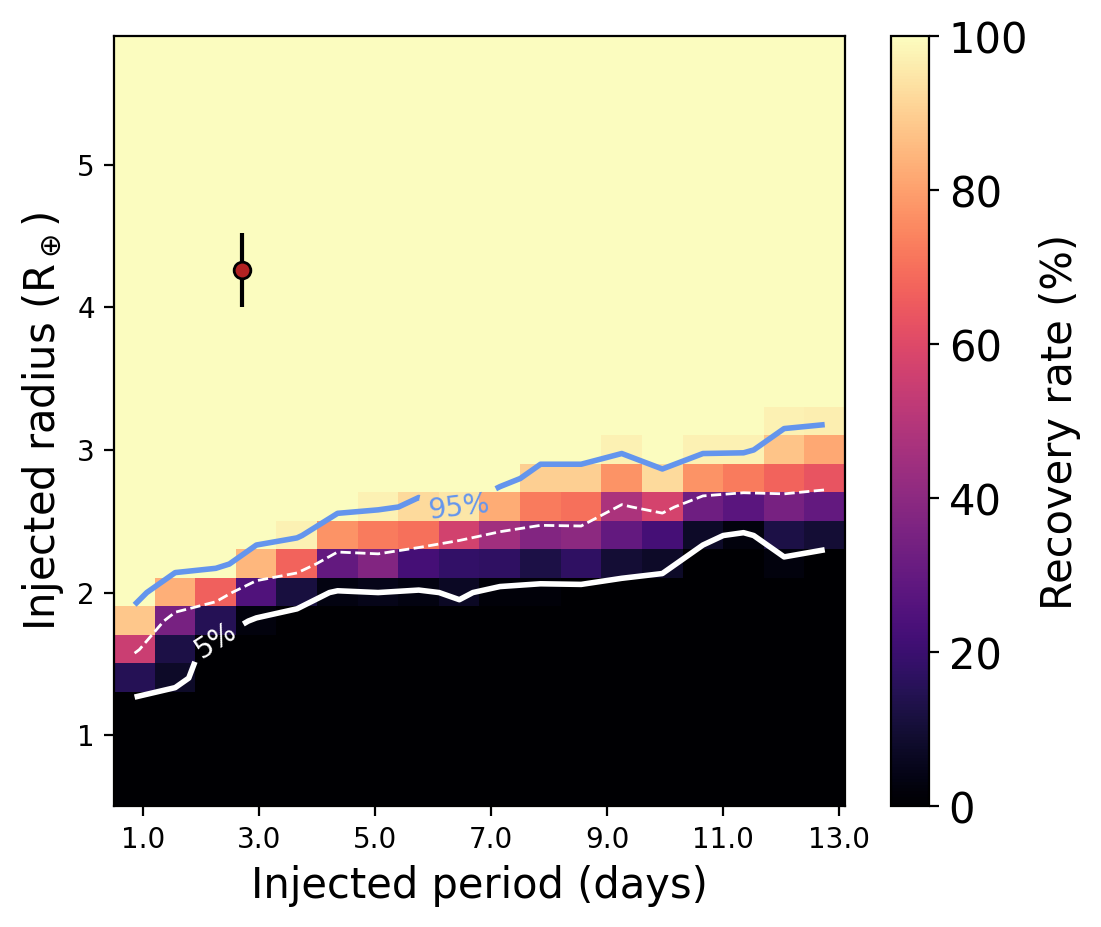}
    \caption{Injection-recovery test performed using the \texttt{MATRIX ToolKit}. The solid lines show the 95\% (top; blue) and 5\% (bottom; white) recovery rates. The dashed white line represents the 50\% recovery rate. The red dot is the position of TOI-2407\,b in period--radius space.}
    \label{fig:IR}
\end{figure}

Even if no evidence of other planetary signals were found, there could be planets in the system that cannot be retrieved with the TESS photometric precision. We used the \texttt{MATRIX ToolKit}\footnote{{\tt tkmatrix:}~\url{https://github.com/PlanetHunters/tkmatrix}} \citep[][]{tkmatrix} to calculate the detection limits of a planet in our TESS data. \texttt{MATRIX} performs transit injection-recovery by injecting synthetic planets into our TESS lightcurves, combining all available sectors. The parameter space explored covered planetary radii from 0.5 to 6\,R$_{\oplus}$  in steps of 0.2\,R$_{\oplus}$, and orbital periods from 0.5 to 13 days in steps of 0.7 days. Each ($R_P$, $P$) pair was tested at four different orbital phases (i.e., $T_0$ values), resulting in a total of 491 scenarios. After injection, \texttt{MATRIX} detrends the light curves using a bi-weight filter with a 0.5 day window. A synthetic planet is considered recovered if its epoch matches the injected epoch within 1 hour and the recovered period is within 5\% of the injected value. For simplicity, all injected planets assume zero eccentricity and impact parameter, meaning the derived detection limits represent an optimistic case \citep[see e.g. ][]{Eisner2020, Pozuelos2020}. The results from the injection-recovery are shown in Figure \ref{fig:IR}.  

\begin{figure}
    \centering\includegraphics[width=0.45\textwidth]{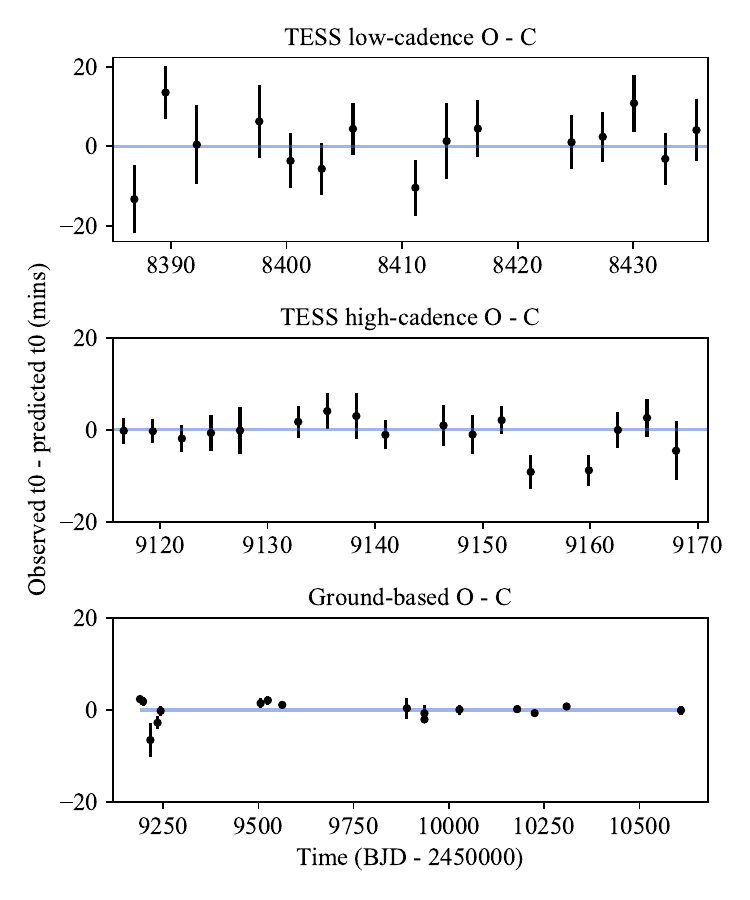}
    \caption{Difference between predicted and observed time for each of the transits to search for transit timing variations (TTVs).}
    \label{fig:TTVs}
\end{figure}

Finally, we looked for the gravitational effect of undetected planets in the system with a TTV search. We fixed our found fit parameters and allowed the mid-transit timing to vary on each observed transit. The difference between observed and predicted is shown in Figure \ref{fig:TTVs}. We found no significant evidence of TTVs. 

\bigskip
\section{{Discussion}}
\label{sec:discuss}

\begin{figure}
    \centering\includegraphics[width=0.95\linewidth,trim=5 5 10 30,clip]{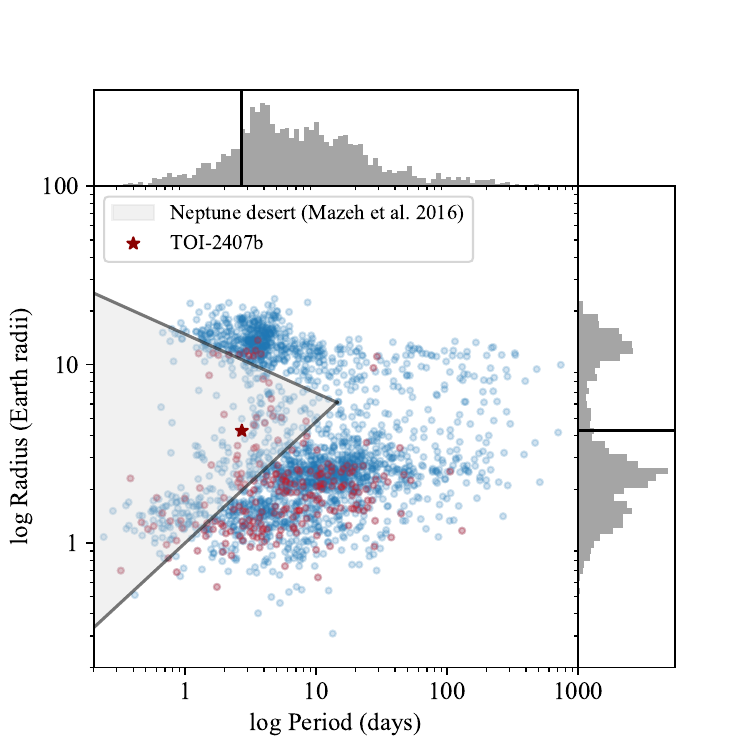}
    \caption{Exoplanets discovered with a radius measurement of at least 10\% precision shown in log$_{10}$ period-radius space. Planets discovered around M-dwarfs are shown in red. TOI-2407\,b is shown as a red star. Data from NASA Exoplanet Archive, January 2025:~\url{https:// exoplanetarchive.ipac.caltech.edu}}
    \label{fig:desert}
\end{figure}

TOI-2407\,b lies in the Neptune desert in period--radius space \citep[][]{Mazeh2016}. This provides an opportunity to test the prevailing theoretical models for the existence of this region, encompassing both formation and atmospheric evolution processes. As shown in Figure \ref{fig:desert}, TOI-2407\,b lies next to the 'Neptune ridge', an overdensity of Neptune-like planets on 3.2--5.7 day period orbits \citep[][]{ridge}. This ridge could arise as a stopping point in their migration path after which planets would undergo photoevaporation. Thus, planets found inside the desert past the ridge could have experienced high-eccentricity migration \citep[][]{Fortney2021} at a late stage, where the star was no longer emitting strongly in the X-ray and XUV. 

The  bounds of the Neptune desert were found using predominantly FGK-type stars. Since M-dwarf planets at a given orbital distance receive lower irradiation than those around earlier-type stars, an alternative way to define the desert is in the irradiation–radius rather than period–radius plane \citep[][]{MacDonald2019, Kanodia2021, Powers2023}. Under this definition, TOI-2407\,b would not be classified as part of the M-dwarf Neptune desert. However, TOI-2407\,b would still present a valuable case for studying photoevaporation in M-dwarf planets, and how it can play a different role in shaping the demographics of such systems compared to FGK stars \citep[seen, e.g.][]{LuquePalle}. A density measurement of TOI-2407\,b will be useful for distinguishing between different formation and evolution models and understanding how the host star influences the desert. A low-density measurement would indicate the presence of a substantial gaseous envelope, suggesting that TOI-2407\,b has retained a large fraction of its primordial atmosphere. 

Using the \cite{massradius} mass--radius relation, we predict a mass of $17\pm2$\,M$_{\oplus}$, placing it in the lower-mass regime of the Neptune desert. Radial velocity measurements of TOI-2407\,b will be needed to characterise the mass and bulk density. This could be obtained with high-precision spectrographs with red and near-infrared wavelength coverage, such as NIRPS \citep[][]{nirps}, which can achieve <10 m/s precision for M-dwarfs. 

For possible JWST observations, based on the predicted mass, we derive a Transmission Spectroscopy Metric (TSM) of 69 and an Emission Spectroscopy Metric (ESM) of 12 using the framework described in \cite{Kempton2018}. The NIRSpec/G395H instrument, which provides higher spectral resolution in the infrared (2.87–-5.14\,$\text{\textmu}$m), would allow for constraints on the presence of carbon-bearing molecules such as CO$_2$, CO and CH$_4$ in the atmosphere, as well as metallicity measurements. C/O ratio measurements with NIRSpec/G395H or through emission spectroscopy with MIRI (5–-12\,$\text{\textmu}$m) could aid in understanding the atmospheric chemistry of TOI-2407\,b and its formation and evolution history. 

TOI-2407\,b also presents a unique opportunity for comparative studies with Gliese 436\,b \citep[][]{gj436}, a well-characterized warm Neptune orbiting an M2.5\,V dwarf with a period of 2.64 days and a radius of 4.33 ± 0.18\,R$_{\oplus}$ \citep[][]{Deming2007}. Gliese 436\,b has been repeatedly observed in transmission and emission spectroscopy, revealing a CH$_4$-deficient, metal-rich atmosphere (Spitzer;  
\citealt{Stevenson2009,  Lanotte2014, Morley2017}; JWST; \citealt{Mukhjerjee2025}). Additionally, atmospheric escape has been observed on Gliese 436\,b through Lyman-$\alpha$ transit observations detecting an extended hydrogen envelope driven by XUV irradiation \citep[][]{Ehrenreich2015}. TOI-2407\,b, with its similar radius and period orbit around an M-dwarf, occupies a comparable parameter space and could provide insight into whether such chemical and dynamical processes are common among Neptunes in this regime. A comparative study with Gliese 436\,b could help constrain potential atmospheric compositions and refine models of atmospheric evolution in the Neptune desert, particularly for M-type stars.  

\section{Conclusions}

This work presents the validation of TOI-2407\,b, a Neptune-sized exoplanet with a radius of $4.26\pm0.26$\,R$_\oplus$ orbiting an M2-type star with a period of 2.7 days. The planet was initially identified through TESS photometry and confirmed via multi-wavelength ground-based transit observations, including data from the near-infrared CMOS detector SPIRIT at the SSO. We report a stellar irradiation of $41\pm4$\,S$_\oplus$, corresponding to a zero-albedo equilibrium temperature of $705\pm12$\,K. The planetary parameters place TOI-2407\,b within the bounds of the Neptune desert in period--radius space, but not in irradiation-radius space.  

Further characterisation of TOI-2407\,b through radial velocity measurements will be necessary to constrain its mass and bulk density, to constrain its composition and evolutionary history. Given its location near the boundary of the period--radius Neptune desert, additional spectroscopic observations could provide insights into the role of atmospheric mass loss and planetary migration in shaping this region of parameter space. 

With an extensive dataset comprising months of SPIRIT observations on a large sample of targets, we will conduct a systematic analysis to further assess the viability of CMOS-based infrared detectors in ground-based exoplanet transit studies. Expanding the sample of transiting exoplanets detected in the infrared will be crucial for improving constraints on planetary demographics around M-dwarfs and for refining models of planet formation and atmospheric evolution in the low-mass stellar regime. SPIRIT will continue to observe the coldest stars, enhancing the SPECULOOS Transit Survey’s ability to detect and characterise planets around ultracool dwarfs.

\section*{Acknowledgements}
Funding for the \tess\ mission is provided by NASA's Science Mission Directorate. We acknowledge the use of public \tess\ data from pipelines at the \tess\ Science Office and at the \tess\ Science Processing Operations Center. This research has made use of the Exoplanet Follow-up Observation Program website, which is operated by the California Institute of Technology, under contract with the National Aeronautics and Space Administration under the Exoplanet Exploration Program. This paper includes data collected by the \tess\ mission that are publicly available from the Mikulski Archive for Space Telescopes (MAST).
Based on data collected by the SPECULOOS-South Observatory at the ESO Paranal Observatory in Chile.The ULiege's contribution to SPECULOOS has received funding from the European Research Council under the European Union's Seventh Framework Programme (FP/2007-2013) (grant Agreement n$^\circ$ 336480/SPECULOOS), from the Balzan Prize and Francqui Foundations, from the Belgian Scientific Research Foundation (F.R.S.-FNRS; grant n$^\circ$ T.0109.20), from the University of Liege, and from the ARC grant for Concerted Research Actions financed by the Wallonia-Brussels Federation. This work is supported by a grant from the Simons Foundation (PI Queloz, grant number 327127).
This research is in part funded by the European Union's Horizon 2020 research and innovation programme (grants agreements n$^{\circ}$ 803193/BEBOP), and from the Science and Technology Facilities Council (STFC; grant n$^\circ$ ST/S00193X/1, and ST/W000385/1).
The material is based upon work supported by NASA under award number 80GSFC21M0002.
We acknowledge funding from the European Research Council under the ERC Grant Agreement n. 337591-ExTrA. 
Based on data collected by the TRAPPIST-South telescope at the ESO La Silla Observatory. TRAPPIST
is funded by the Belgian Fund for Scientific Research (Fond National de la Recherche Scientifique, FNRS) under the grant PDR T.0120.21, with the participation of the Swiss National Science Fundation (SNF). E.J. and M.G. are FNRS Research Directors.
The postdoctoral fellowship of KB is funded by F.R.S.-FNRS grant T.0109.20 and by the Francqui Foundation. This publication benefits from the support of the French Community of Belgium in the context of the FRIA Doctoral Grant awarded to M.T. BVR thanks the Heising-Simons Foundation for support. F.J.P acknowledges financial support from the grant CEX2021-001131-S funded by MCIN/AEI/10.13039/501100011033and through projects PID2019-109522GB-C52 and PID2022-137241NB-C43. KAC acknowledges support from the TESS mission via subaward s3449 from MIT. B.-O. D. acknowledges support from the Swiss State Secretariat for Education, Research and Innovation (SERI) under contract number MB22.00046. E. J. is a Belgian FNRS Senior Research Associate. YGMC acknowledges support from UNAM-PAPIIT-IG101321.
This material is based upon work supported by the National Aeronautics and Space Administration under Agreement No.\ 80NSSC21K0593 for the program ``Alien Earths''.
The results reported herein benefited from collaborations and/or information exchange within NASA’s Nexus for Exoplanet System Science (NExSS) research coordination network sponsored by NASA’s Science Mission Directorate.
Based in part on observations obtained at the Southern Astrophysical Research (SOAR) telescope, which is a joint project of the Minist\'{e}rio da Ci\^{e}ncia, Tecnologia e Inova\c{c}\~{o}es (MCTI/LNA) do Brasil, the US National Science Foundation’s NOIRLab, the University of North Carolina at Chapel Hill (UNC), and Michigan State University (MSU).
Based in part on observations obtained through the Astronomical Event Observatory Network (AEON), a joint endeavor of the Las Cumbres Observatory and of NSF NOIRLab, which is managed by the Association of Universities for Research in Astronomy (AURA) under a cooperative agreement with the U.S. National Science Foundation.
This work makes use of observations from the LCOGT network. Part of the LCOGT telescope time was granted by NOIRLab through the Mid-Scale Innovations Program (MSIP). MSIP is funded by NSF.
Resources supporting this work were provided by the NASA High-End Computing (HEC) Program through the NASA Advanced Supercomputing (NAS) Division at Ames Research Center for the production of the SPOC data products.
CJM thanks J. Dhandha for the comments and feedback on the draft. This work was supported by the University of Cambridge Harding Distinguished Postgraduate Scholars Programme.

\section*{Data Availability}

TESS and ground-based data available on ExoFOP-TESS: \url{https://exofop.ipac.caltech.edu/tess/target.php?id=153078576}.



\bibliographystyle{mnras}
\bibliography{refs}



\newpage
\appendix

\section{Limb darkening and detrending parameters}
\label{limbdarkening}
\begin{table}
\centering
\renewcommand{\arraystretch}{1.2} 
\begin{tabularx}{0.5\textwidth}{l>{\centering}X>{\centering\arraybackslash}c}
\toprule
\textbf{Parameter} & \textbf{Value} & \textbf{Priors} \\ \midrule
\multicolumn{3}{l}{Limb-darkening coefficients} \\ \midrule
\( q_{1,\text{TESS}} \) & \(0.21^{+0.22}_{-0.12}\,\) & \(\mathcal{U}(0, 1)\) \\
\( q_{2,\text{TESS}} \) & \(0.22^{+0.19}_{-0.16}\,\) & \(\mathcal{U}(0, 1)\) \\
\( q_{1,\text{SPIRIT}} \) & \(0.60^{+0.23}_{-0.22}\,\) & \(\mathcal{U}(0, 1)\) \\
\( q_{2,\text{SPIRIT}} \) & \(0.26^{+0.16}_{-0.15}\,\) & \(\mathcal{U}(0, 1)\) \\
\( q_{1,\text{g}'} \) & \(0.48^{+0.31}_{-0.25}\,\) & \(\mathcal{U}(0, 1)\) \\
\( q_{2,\text{g}'} \) & \(0.24^{+0.17}_{-0.16}\,\) & \(\mathcal{U}(0, 1)\) \\
\( q_{1,\text{z}} \) & \(0.52^{+0.29}_{-0.26}\,\) & \(\mathcal{U}(0, 1)\) \\
\( q_{2,\text{z}} \) & \(0.27^{+0.17}_{-0.18}\,\) & \(\mathcal{U}(0, 1)\) \\
\( q_{1,\text{V}} \) & \(0.34^{+0.29}_{-0.21}\,\) & \(\mathcal{U}(0, 1)\) \\
\( q_{2,\text{V}} \) & \(0.15^{+0.17}_{-0.11}\,\) & \(\mathcal{U}(0, 1)\) \\
\( q_{1,\text{I+z}} \) & \(0.33^{+0.28}_{-0.21}\,\) & \(\mathcal{U}(0, 1)\) \\
\( q_{2,\text{I+z}} \) & \(0.26^{+0.15}_{-0.16}\,\) & \(\mathcal{U}(0, 1)\) \\
\( q_{1,\text{LCO}} \) & \(0.34^{+0.33}_{-0.24}\,\) & \(\mathcal{U}(0, 1)\) \\
\( q_{2,\text{LCO}} \) & \(0.31^{+0.13}_{-0.19}\,\) & \(\mathcal{U}(0, 1)\) \\
\( q_{1,\text{ExTrA}} \) & \(0.06^{+0.13}_{-0.04}\,\) & \(\mathcal{U}(0, 1)\) \\
\( q_{2,\text{ExTrA}} \) & \(0.21^{+0.19}_{-0.16}\,\) & \(\mathcal{U}(0, 1)\) \\
\( u_{1,\text{TESS}} \) & \(0.19^{+0.14}_{-0.14}\,\) & - \\
\( u_{2,\text{TESS}} \) & \(0.28^{+0.23}_{-0.21}\,\) & - \\
\( u_{1,\text{SPIRIT}} \) & \(0.38^{+0.20}_{-0.21}\,\) & - \\
\( u_{2,\text{SPIRIT}} \) & \(0.38^{+0.26}_{-0.27}\,\) & - \\
\( u_{1,\text{g}'} \) & \(0.32^{+0.23}_{-0.23}\,\) & - \\
\( u_{2,\text{g}'} \) & \(0.35^{+0.26}_{-0.25}\,\) & - \\
\( u_{1,\text{z}} \) & \(0.35^{+0.23}_{-0.24}\,\) & - \\
\( u_{2,\text{z}} \) & \(0.34^{+0.28}_{-0.27}\,\) & - \\
\( u_{1,\text{V}} \) & \(0.19^{+0.16}_{-0.15}\,\) & - \\
\( u_{2,\text{V}} \) & \(0.39^{+0.24}_{-0.24}\,\) & - \\
\( u_{1,\text{I+z}} \) & \(0.27^{+0.16}_{-0.17}\,\) & - \\
\( u_{2,\text{I+z}} \) & \(0.29^{+0.24}_{-0.23}\,\) & - \\
\( u_{1,\text{LCO}} \) & \(0.33^{+0.25}_{-0.24}\,\) & - \\
\( u_{2,\text{LCO}} \) & \(0.23^{+0.18}_{-0.17}\,\) & - \\
\( u_{1,\text{ExTrA}} \) & \(0.11^{+0.078}_{-0.077}\,\) & - \\
\( u_{2,\text{ExTrA}} \) & \(0.18^{+0.15}_{-0.15}\,\)
\end{tabularx}
\end{table}

\bsp	
\label{lastpage}
\end{document}